\documentclass[10pt,letterpaper,single column]{article} 

\usepackage{ol2}
\usepackage[draft]{hyperref}
\usepackage{amsmath}

\begin{document}


\title{Single beam atom sorting machine}


\author{M. McGovern, T. Gr\"{u}nzweig, A. J. Hilliard and M. F. Andersen$^{*}$}

\address{Jack Dodd Centre for Quantum Technology, Department of Physics, University of Otago, New Zealand

$^*$Corresponding author:  mikkel@physics.otago.ac.nz}

\begin{abstract}We create two overlapping one-dimensional optical lattices using a single laser beam, a spatial light modulator and a high numerical aperture lens. These lattices have the potential to trap single atoms, and using the dynamic capabilities of the spatial light modulator may shift and sort atoms to a minimum atom-atom separation of $1.52~\mu$m. We show how a simple feedback circuit can compensate for the spatial light modulator's intensity modulation.
\end{abstract}

\section{Introduction}
Individual neutral atoms trapped in optical dipole traps present a feasible approach to the construction of a quantum computer \cite{Yukalov2009a,Wineland2011}, as well as providing a versatile platform for direct investigation of  quantum phenomena \cite{Yukalov2009b,Oztop2009,Zenesini2010}. The capability to manipulate and rearrange the relative position of individual atoms in the optical dipole traps is crucial to the above fields, and has been the subject of intense research. Different approaches to performing this manipulation include the use of acoustic optic modulators (AOM) to create an `atom sorting machine' \cite{Miroshnychenko2006}, using a spatial light modulator (SLM) to create dynamic atom traps \cite{Bergamini2004,He2010,Brandt2011} and using piezo-controlled mirrors \cite{Beugnon2007} to change the position of trapped atoms.

In particular, the atom sorting machine \cite{Miroshnychenko2006} represented an important step in scientists' ability to manipulate the microscopic world. In this setup, atoms are trapped in the antinodes of two 1-D crossed optical standing waves \cite{Yukalov2009}. The atom trapping antinodes are shifted through the use of AOMs, and are able to shift and sort trapped neutral atoms to a distance of 10~$\mu$m of each other. The interatomic distance of 10~$\mu$m is limited by the size of the beam waist of each of the overlapping 1-D optical standing waves. The atom sorting machine, combined with the recently demonstrated neutral atom Rydberg gate \cite{Gaetan2009,Urban2009}, are key components towards the development of a neutral atom based quantum computer \cite{Saffman2005}. However, as shown in Refs.~\cite{Gaetan2009,Urban2009}, $\sim4~\mu$m is the interatomic distance needed in order to create the neutral atom Rydberg gate. Introducing high numerical aperture lenses into the experiment described in Ref. \cite{Miroshnychenko2006} could decrease the waists of the overlapping 1-D optical standing waves, and therefore the interatomic distance by up to an order of magnitude, but 4 lenses would be needed to achieve this, severely restricting the optical access in such a setup.

Holograms, on the other hand, have the potential to create arbitrarily shaped optical dipole traps. Therefore one could create two 1-D lattices with the interatomic distance required for a Rydberg gate, with holograms, laser light and an appropriate single lens. Static and dynamical holograms can be created with spatial light modulators \cite{Dufresne1998}. The use of certain SLM modules can introduce an intensity fluctuation to a diffracted light beam \cite{Horst2008}. A fluctuating trapping beam intensity shifts a trapped atom's resonances \cite{Grimm2000}, and therefore it is difficult to resonantly address such an atom. Furthermore, the resulting time dependent trapping potential can lead to heating and possibly loss of the trapped atom.

We recently demonstrated a high-efficiency method of loading individual atoms to an optical micro-trap \cite{Grunzweig2010}. In that experiment a large number of atoms are first loaded to a microscopic optical trap, whereupon they are irradiated with laser light to induce light-assisted collisions. The energy gained through these collision allow atoms to escape the trap, until only one atom remains trapped. Our demonstrated  loading efficiency of  83\% allows for   scaling the number of singly occupied  micro-traps to beyond a few, but sorting is still required for larger systems. We used   blue-detuned light for fluorescence detection, together with a   variant of  Sisyphus cooling to image the trapped atoms \cite{McGovern2011}.

In this paper we demonstrate how we use analytical holograms produced with an SLM, and a single high numerical aperture lens, to create two 1D crossed optical standing waves. Here the antinodes of the standing waves can be used as a single beam atom sorting machine capable of achieving an interatomic separation of 1.52~$\mu$m.  We show how with the use of a simple feedback loop, the undesirable intensity noise created by the SLM can be decreased by 90~$\%$.

\section{Setup}
\subsection{Trapping atoms with light}
\begin{figure}
\begin{center}
  \includegraphics[width=8.0cm]{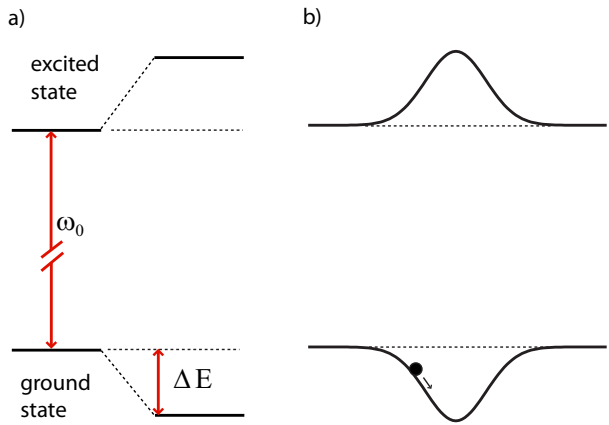}
  \caption{a) The effect of far off resonant light on the energy levels of a two level atom. b) The  spatially dependent energy levels resulting from a light field from a Gaussian beam in which the atom sees the ground state as a potential well, and may become trapped if its energy is small enough.}
  \label{dipole}
  \end{center}
\end{figure}

Neutral atoms can be spatially confined in red-detuned optical dipole traps. When a two level atom is in a spatially dependent far off resonant light field, its ground and excited state energy levels experience a light shift $\Delta E$ equal to \cite{Grimm2000}:
\begin{equation}
\Delta E = \pm \frac{3 \pi c^{2}}{2 \omega_{0}^{3} } \frac{\Gamma}{\Delta}I(\textbf{r}) \label{Eqn_Light_Shift}
\end{equation}
where $\omega_{0}, \Gamma, \Delta, I(\textbf{r})$ are the transition angular frequency of the two level atom, the decay rate of the excited atom, the detuning between the light angular frequency and atomic transition angular frequency, and the spatially dependent intensity of the light field respectively. The $\pm$ represents whether the energy level has been shifted up or down. For far off resonance red detuned light the ground state energy level is shifted down, and the excited energy level is shifted up. A spatially dependent field, such as that of a Gaussian beam, produces a ground state potential well, $U(\textbf{r}) =  \Delta E_{-}(\textbf{r})$, in which an atom can be trapped.
 The potential of this atom trap is proportional to the local intensity of the laser beam, as in the schematic in Fig.~\ref{dipole}b. Such optical dipole traps can be formed in various ways, for example as the focus of a tightly focused laser beam \cite{Schlosser2001}, or arrays of dipole traps can be formed as the antinodes of a standing wave of two counter-propagating laser beams \cite{Miroshnychenko2006}.

\subsection{Setup}

\begin{figure}
\begin{center}
  \includegraphics[width=16.5cm]{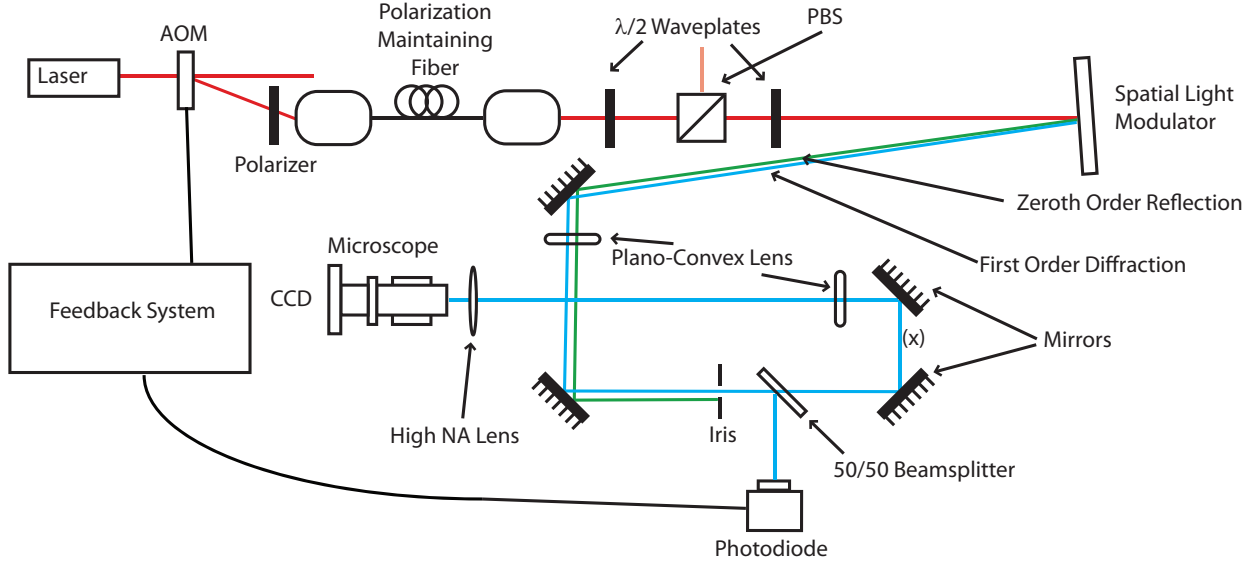}
  \caption{Schematic of the setup. Light projected onto the SLM is diffracted and spatially filtered before being imaged through a high numerical aperture lens. The resulting image is viewed using the microscope. The point ``(x)'' is the space where a $\frac{\lambda}{2}$ waveplate and PBS is added to calibrate the SLM. To compensate for the intensity flicker of the SLM, the power of a small part of the diffracted beam is measured, passed through an electronic feedback system, and fed back into the AOM.}
  \label{setup}
  \end{center}
\end{figure}

Figure~\ref{setup} is a schematic of the setup we built to investigate the use of an SLM as a single beam atom sorting machine with a sub-micron lattice constant. A laser beam red-detuned from 780~nm is passed through an AOM, the first order diffracted beam is passed through a polarizer and coupled into a polarization maintaining fiber. The power in the beam is controlled through the AOM driver. To trap neutral atoms, one generally needs a strong intensity gradient, and therefore a high intensity light field \cite{Grunzweig2010}. However, here we use a very low power beam during investigations in order to not saturate the imaging camera. The outcoupled beam from the fiber has a Gaussian profile with waist of 5~mm, and power of 15~nW. To ensure a pure polarisation, we
pass this beam through a $\frac{\lambda}{2}$ waveplate and polarizing beamsplitter (PBS). A subsequent $\frac{\lambda}{2}$ waveplate is used to orientate the polarization of light for the SLM. We use a Holoeye Pluto Phase Only Reflective Modulator. The SLM face is 16.6 $\times$ 10.2~mm, with 1920 $\times$ 1080 8~$\mu$m pixels. The SLM takes its input signal from the green color channel of the graphics card in the computer to which it is connected. The SLM can respond to 256 gray-scale levels displayed on the computer monitor, and therefore each pixel can impart 256 different phase levels to an incident laser beam. The SLM has a refresh rate of 60~Hz.
The SLM is tilted at a small angle to direct the reflection and diffracted beams away from the input optics.
We use a blazed hologram to diffract the incident beam, and its first order is used to create the atom-sorting machine. We image the diffracted beam with a two lens, 1$\times$ telescope configuration. The reflected beam is spatially filtered with an iris at the focus of both 300~mm plano-convex lenses within the telescope, where the reflected and diffracted beams are well separated (Fig.~\ref{setup}).
A beamsplitter directs a portion of the spatially filtered beam onto a photodiode for use in a feedback system to minimize the intensity flicker produced by the refresh rate of the SLM, as is described below. The spatially filtered beam is then focused through a high numerical aperture lens (NA = 0.55), identical to the one used for single atom trapping in Ref. \cite{Grunzweig2010}. The light structures produced are imaged with a 100x magnification microscope, which has a NA = 0.7 objective lens, onto a charge coupled device (CCD) camera.

\subsection{SLM calibration}
We use blazed holograms on the SLM to form the dipole trap structures. The main advantage of this approach is the high diffraction efficiency one can obtain (up to 83$\%$ \cite{HOLOEYE}). To achieve high diffraction efficiency, the phase response of the SLM must be calibrated for use in a given experiment. A linear response between the SLM's 256 gray-scale inputs and a $0-2\pi$ phase shift is desired. But the phase added to incident light is  dependant on the wavelength of the laser and the angle of the SLM to the incoming beam. The SLM has a Look Up Table (LUT) that stores the input-grayscale-
to-phase-conversion. To calibrate the LUT to the input angle and light
wavelength used we used a method inspired from Ref.  \cite{Wilson2007} in which we interfere a
beam phase-modulated by the SLM, with a beam that receives no phase-modulation from the SLM. We begin with a linearly polarized beam traveling in the $\textbf{\^z}$ direction, in complex notation its electric field is given  by:
\begin{center}
\begin{equation}
\textbf{E}(z,t) = \frac{1}{\sqrt{2}}\textit{E}_{0}\left(\textbf{\^{x}}+ \textbf{\^{y}}\right) \exp(ikz-i\omega t),
\end{equation}
\end{center}
where $E_0$ is the field amplitude, $\omega$ is the angular frequency, $k$ is the wave number and  $\textbf{\^{x}}, \textbf{\^{y}}$  are the unit polarization vectors.
The SLM will only phase modulate incident light of a certain polarization. Here the modulation axis is the \textbf{\^{x}} axis. Therefore after being reflected by  the SLM, the electric field of the beam is:
\begin{center}
\begin{equation}
\textbf{E}(z,t) = \frac{1}{\sqrt{2}}\textit{E}_{0}\left(\textbf{\^{x}}\exp(i\theta)  + \textbf{\^{y}} \right)  \exp(ikz-i\omega t)
\end{equation}
\end{center}
where $\theta$ is the phase added to the component of the beam with \textbf{\^{x}} polarization. $\theta$ is dependant on the LUT value for that particular gray scale level. Passing this beam through a $\frac{\lambda}{2}$ waveplate, with fast axis at an angle of $\frac{\pi}{8}$ with respect to the $\textbf{\^{x}}$ polarization axis, creates the following electric field:
\begin{center}
\begin{eqnarray}
\textbf{E}(z,t) &=& \frac{1}{2}\textit{E}_{0}\left(\exp(i\theta)+ 1\right)\textbf{\^{x}} \exp(i k z-i\omega t) \\
 &+& \frac{1}{2}\textit{E}_{0}\left(\exp(i\theta)- 1\right))\textbf{\^{y}}   \exp(i k z-i\omega t).
\end{eqnarray}
\end{center}
This beam passes through a PBS and the intensity from the output arm that reflects the component of light of $\textbf{\^{x}} $ polarization is:
\begin{equation}
I(\theta) = |E_{x}|^{2} = |\textit{E}_{0}|^{2}\cos^{2}\left(\frac{\theta}{2}\right). \label{eqn_intensity}
\end{equation}
We play a ``movie'' on the SLM: a sequence of 256 frames, corresponding to the 256 available gray scale levels, at a rate of 4 frames per second, and in each frame all pixels identically display one of the 256 levels available to them. We record the intensity of light, $I(\theta)$, in one of the output arms of the PBS. This measured intensity changes as a result of each frame of the movie changing, and from Eq.~\ref{eqn_intensity} one can deduce the phase added for each particular gray scale level associated with each frame. From these intensity measurements a new LUT is created that will produce a linear phase response and this LUT is written to the SLM.

This simple inline method means that to calibrate the SLM, we only need to place a $\frac{\lambda}{2}$ waveplate, a PBS and a photodetector into the beam reflected from the SLM, at the point marked ``(x)'' in the existing setup in Fig.~\ref{setup}. The method is convenient as it calibrates the SLM \emph{in situ} for the correct angle and wavelength it is intended to be used for within its future experiments. Changing the incident angle or wavelength of the laser beam will decrease the diffraction efficiency of the SLM. The optical components used for calibration can be taken out of the setup very easily, without affecting the beam incident angle on the SLM. After this calibration we can diffract 79$\%$ of incoming light for 780~nm, close to the maximum $83\%$ possible \cite{HOLOEYE} for the default wavelength of 633~nm.

\subsection{Minimizing SLM flicker}
\begin{figure}
\begin{center}
  \includegraphics[width=8.0cm]{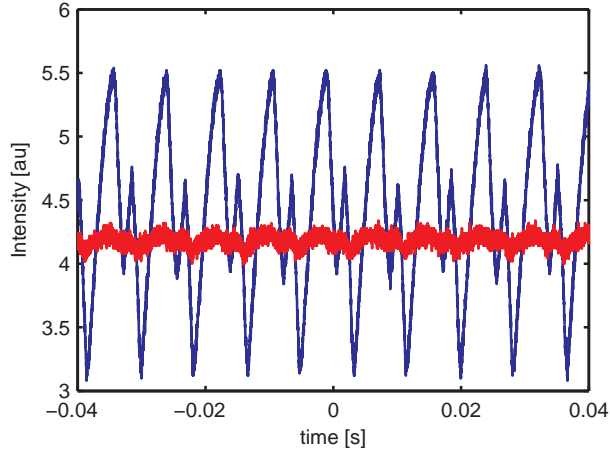}
  \caption{Measured intensity (blue line)  of first order diffracted beam from SLM. A feedback circuit is switched on an reduces the intensity flicker (red line) by $90\%$.}
  \label{feedback}
  \end{center}
\end{figure}

The Holoeye SLM used in this experiment has a refresh rate of 60~Hz. The Liquid Crystal (LC) pixels themselves are addressed at the addressing rate of 120~Hz. Due to the binary nature of the addressing scheme, the liquid crystals are continuously moving, contributing to a phase shift error at the addressing rate \cite{Kakarenko2010}. This adds a flicker to the intensity of the diffracted beam of up to $\pm28\%$. The measured intensity is displayed as the blue line in  Fig.~\ref{feedback}. When trapping atoms in optical dipole traps formed from an SLM, an intensity modulation will lead to a corresponding modulation of the trapping potential and the atomic resonances ($U(\textbf{r}) \propto I $), making such trapped atoms difficult to probe and investigate \cite{Horst2008}. To minimize this flickering effect we direct a small portion of the first order beam from the SLM onto a photodiode, and the resulting signal passes through a simple electronic feedback loop, to the AOM driver, as in Fig.~\ref{setup}. The red line in Fig.~\ref{feedback} is the resulting measured intensity, with the amplitude of the noise reduced by 90\%. Instead of a feedback circuit, one could feed the correct waveform to the AOM in order to completely cancel the flicker. However in our experiments we will dynamically change the holograms on the SLM, and with different holograms, different intensity modulations exist, a real time feedback circuit is needed to minimize the intensity modulations.


\section{Dipole Traps with High NA lens and SLM}

\subsection{Static Dipole Traps}
Using this setup, a single dipole trap can be formed by having a blazed grating cover the entire SLM: the diffracted beam retains the incoming beam's spatial profile. Once this is focused through the high NA lens, it creates a single dipole trap with a full-width-half-maximum (FWHM) size of ($0.76\pm 0.04~\mu$m).

\begin{figure}
\begin{center}
  \includegraphics[width=8.0cm]{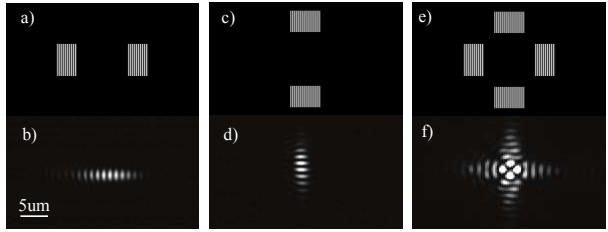}
  \caption{a) Gray scale image that produces a hologram on the SLM. Here black corresponds to a 0 phase shift, and white a $2\pi$ phase shift.  b) Horizontal lattice pattern produced with above gray scale image. c) Gray scale image and d) vertical lattice. e) This gray scale image produces f) horizontal and vertical lattices overlapped. Notice the interference pattern where the two lattices have interfered.}
  \label{lattice}
  \end{center}
\end{figure}
Figures~\ref{lattice}a-f contain images of arrays of optical dipole traps, as well as the gray scale images, corresponding to the holograms formed on the SLM that produce them. The gray-scale images in Figs.~\ref{lattice}a, c and e are made up of pairs of rectangles, $200 \times 300$~pixels, separated by 500~pixels containing a vertical blazed grating. The blazed grating has a period of 20~pixels. A pixel on the SLM, whose corresponding pixel from the gray-scale image is black, will add a 0 radian phase shift to incident light. Similarly white corresponds to a $2\pi$ phase shift.
In the case of producing the array of traps in Fig.~\ref{lattice}a, two sections of the beam incident on the SLM are diffracted, corresponding to the light incident on the two sections of blazed grating in the hologram.  The non-diffracted part of the beam is spatially filtered, and the diffracted parts of the beam are focused by the high NA lens. These two parts of the diffracted light beam will interfere once they are focused by the high NA lens to form the standing wave in the horizontal direction in Fig.~\ref{lattice}b.
By changing the layout of the gray-scale image, and therefore the hologram, one can produce an array of dipole traps in the vertical direction, as in Fig.~\ref{lattice}d. The lattice constant can also be changed by changing the distance between the two blazed sections of the SLM hologram. We have obtained a sub-micron lattice constant of $0.76\pm 0.04 \mu$m, and this is limited by the numerical aperture of the aspheric lens. The error in this measurement is due to the pixel size of the CCD camera. The minimum transverse waist of the lattices in Fig.~\ref{lattice} for this setup is $1.01\pm 0.04\mu$m (the full-widthhalf-
maximum of the horizontal lattice in the vertical direction). This waist defines how many traps will overlap
(2 in this case) if we project both a lattice in the vertical and horizontal direction on top of each other as in Fig~\ref{lattice}f. This ultimately limits the atom-atom separation that we can obtain when using this setup to rearrange and sort single atoms. The analytical holograms were created using MATLAB$^{\textregistered}$ software, requiring less processing time than creating holograms by taking numerical fourier transforms of desired light patterns.

\subsection{Dynamic optical lattices for atom sorting}
Trapped neutral atoms will remain in the moving dipole traps, if these traps move adiabatically \cite{Miroshnychenko2006,Kuhr2001,Cizmar2011}. We create moving dipole traps by dynamically changing the holograms that produced the static dipole traps above. If a phase shift is introduced into one of the blazed gratings in the gray-scale image in Fig.~\ref{lattice}a, the resulting array of dipole traps will be shifted along the lattice. However, the overall position of the entire array will not move, only the relative position of the dipole traps within the array envelope. A relative phase change of $2\pi$ between the blazed gratings results in the movement of the lattice of a distance of one lattice constant. Therefore by incremental changing the relative phase of one of the blazed gratings in Fig.~\ref{lattice}a, we produce a horizontal conveyor belt capable of moving atoms. The conveyor belt can drive in either direction, by changing the direction of relative phase shift. Similarly shifting atoms up and down can be achieved in a vertical conveyor belt by changing the phase of one of the blazed gratings in Fig.~\ref{lattice}c relative to the other.

The atom-sorting machine described in Ref.~\cite{Miroshnychenko2006} sorts atoms in both order and position by having an overlapping vertical and horizontal lattice pattern that can move independently of each other. By using the vertical conveyor belt to extract atoms from the horizontal lattice, followed by a shift of the horizontal lattice and a reinsertion, atoms can be repositioned in the horizontal lattice pattern. This can be reproduced using an SLM by applying a hologram similar to that in Fig.~\ref{lattice}e where the relative phase is shifted between the left and right blazed gratings to drive the horizontal conveyor belt. Independently shifting the relative phase between the top and bottom blazed grating will drive the vertical conveyor belt. However, following this procedure, a complex interference pattern is produced in the overlap of the two lattices. An atom will be lost when attempting to transport it through this central interference. To overcome this, when driving the horizontal conveyor belt, we change the absolute phase of both the left and right blazed gratings in the hologram. Specifically, an incremental phase is added to the left blazed grating, whilst simultaneously being subtracted from the right blazed grating; the phase of the top and bottom gratings are held constant. This is as we would expect for creating an interfering moving lattice system: in order to maintain constructive interference between the antinodes of the vertical and the moving horizontal lattice, the relative phase between horizontal antinodes and vertical antinodes must not change. A similar procedure is used when operating the vertical conveyor belt.

%

\begin{figure}
  \begin{center}
  \includegraphics[width=8.0cm]{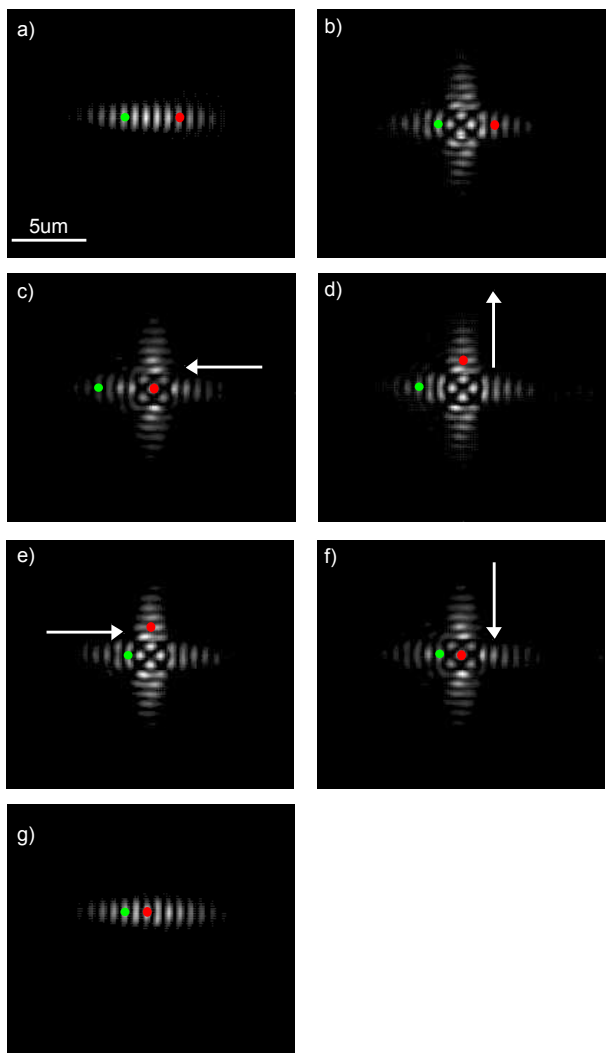}
  \caption{Schematic of how single beam atom-sorting machine could work. a) Two atoms may be loaded into horizontal lattice and needed to be moved closer together. b) Vertical lattice turned on. c) One atom moved to center. d) Atom extracted. e) Second atom repositioned. f) Atom reinserted. g) Vertical lattice turned off.}
  \label{atom_sorting}
  \end{center}
\end{figure}

Figure~\ref{atom_sorting} shows a series of images of optical dipole traps created with the SLM, and with colored spots representing atoms superimposed into the images, this illustrates conveyor belt protocol. Figure~\ref{atom_sorting}a shows where two atoms may originally be loaded into the horizontal optical lattice. In Fig.~\ref{atom_sorting} b-d the vertical lattice is turned on, one of the atoms is moved to the center of the horizontal optical lattice using the horizontal optical conveyor belt, then extracted using the optical conveyor belt in the vertical direction. In Fig.~\ref{atom_sorting} e-g the horizontal optical lattice is then moved, the atom is reinserted and the vertical optical lattice is turned off. This shows how atoms could be sorted to within 1.52~$\mu$m (two lattice constants) of each other.

Our scheme to sort atoms has both advantages and disadvantages over the previously demonstrated atom sorting machine \cite{Miroshnychenko2006}. The amount of overlap between horizontal and vertical lattice patterns due to the size of the waist of the lattice patterns ultimately limits the minimum achievable distance between atoms after sorting. For example in \cite{Miroshnychenko2006}, the minimum distance between sorted atoms is $10\mu$m. Because we use a high NA lens, we are able to achieve relatively small waists in the lattice patterns. Our measured minimum lattice waist is $1.01\pm 0.04\mu$m, which when overlapped by a perpendicular lattice pattern is a minimum overlap of two lattice constants ($1.52\pm 0.08 \mu$m). This is
the minimum atom-atom spacing achievable after sorting with this setup, comparing favorable to the interatomic distance
of $\sim4~\mu$m needed in order to create the neutral atom Rydberg gate \cite{Gaetan2009,Urban2009}. However the number of lattice sites in \cite{Miroshnychenko2006} is limited only by the size of their vacuum chamber and wavelength of the lattice pattern. Diffraction aberrations limited our setup to the creation of up to a 1-dimensional lattice with 28 resolvable
sites. However, less than 28 atoms could be ``sorted'' in this case;  as atoms originally trapped on the edges of the lattice will be lost as ``the conveyor belt'' moves to shift atoms.

\section{Conclusions}
We have used a spatial light modulator, a high numerical aperture lens and a single laser beam to create dynamical sub-micron dipole traps which may be used to trap neutral atoms and with the potential to manipulate the atom's spatial position. We can create an overlapping horizontal and vertical lattice of atom trapping potentials with a sub-micron lattice constant. By dynamically changing the SLM hologram we can change the position of the lattice potentials, thereby creating a single beam scheme that may be used for moving and sorting individual atoms, with a potential minimum atom-atom separation of $1.52\mu$m. The intensity `flicker' (a product of the digital nature of the SLM) is compensated with a simple feedback circuit and an acousto-optic modulator.

\section{Acknowledgements}
This work is supported by the New Zealand Foundation for Research, Science and Technology (NZ-FRST) Contract No. NERF-UOOX0703 and a University of Otago Research Grant.

\end{document}